\documentclass[proof]{pasj01}
\usepackage{bm}
\usepackage{color}
\draft 
\Received{$\langle$reception date$\rangle$}
\Accepted{$\langle$acception date$\rangle$}
\Published{$\langle$publication date$\rangle$}
\begin{document}

\title{Developments of Multi-wavelength Spectro-Polarimeter on the Domeless Solar Telescope at Hida Observatory}
\author{
Tetsu \textsc{Anan}\altaffilmark{1, 2},
Yu Wei \textsc{Huang}\altaffilmark{1},
Yoshikazu \textsc{Nakatani}\altaffilmark{1},
Kiyoshi \textsc{Ichimoto}\altaffilmark{1, 3},
Satoru \textsc{Ueno}\altaffilmark{1},
Goichi \textsc{Kimura}\altaffilmark{1},
Shota \textsc{Ninomiya}\altaffilmark{1},
Sanetaka \textsc{Okada}\altaffilmark{1},
Naoki \textsc{Kaneda}\altaffilmark{1}
}
\altaffiltext{1}{ Kwasan and Hida Observatory, Kyoto University, Kamitakara, Gifu 506-1314, Japan }
\altaffiltext{2}{ National Solar Observatory, 8 Kiopa'a Street, Pukalani, HI 96768, USA }
\altaffiltext{3}{ National Astronomical Observatory of Japan, 2-21-1 Osawa, Mitaka, Tokyo 181-8588, Japan }
\email{tanan@nso.edu}


\maketitle

\begin{abstract}
%現在形で
To obtain full Stokes spectra in multi-wavelength windows simultaneously, we developed a new spectro-polarimeter on the Domeless Solar Telescope at Hida Observatory. The new polarimeter consists of a 60 cm aperture vacuum telescope on an altazimuth mount, an image rotator, a high dispersion spectrograph, polarization modulator and analyzer composed of a continuously rotating waveplate with a retardation nearly constant around 127$^{\circ}$ in 500 - 1100 nm and a polarizing beam splitter located closely behind the focus of the telescope, fast and large format CMOS cameras and an infrared camera. The slit spectrograph allows us to obtain spectra in as many wavelength windows as the number of cameras. We characterized the instrumental polarization of the entire system and established the polarization calibration procedure. The cross-talks among the Stokes Q,U and V are evaluated to be about 0.06\% $\sim$ 1.2\% depending on the degree of the intrinsic polarizations. In a typical observing setup, a sensitivity of 0.03\% can be achieved in 20 - 60 second for 500 nm - 1100 nm. The new polarimeter is expected to provide a powerful tool to diagnose the 3D magnetic field and other vector physical quantities in the solar atmosphere.
\end{abstract}

%%%%%%%%%%%%%%%%%%%%%%%%%%%%%%%%%%%%%%%%%%%%%
%%%%%%%%%%%%%%%%%%%%%%%%%%%%%%%%%%%%%%%%%%%%%
%%%%%%%%%%%%%%%%%%%%%%%%%%%%%%%%%%%%%%%%%%%%%
\section{Introduction}
\label{sec.intro}

Since polarization of spectral lines is created from anisotropic physical state in the light source or in the medium through which the light propagates \citep{landi04}, polarimetry makes it possible to diagnose vector physical quantities such as magnetic fields \citep{zeeman97, hanle24}, electric fields \citep{stark13, casini05}, radiation field and stream of particles.
Diagnosis of such physical quantities and, in particular, the determination of three dimensional structure of magnetic field are of crucial importance for the next generation of solar physics \citep{foukal91, stenflo17}. 
However, it is difficult to distinguish the nature of anisotropies through diagnosis made by the polarimetry in a single spectral line.
In addition, single-line polarimetry does not allow us to determine magnetic structures extending from solar photosphere into upper chromosphere, because the polarization in a single line is sensitive to the magnetic field in rather limited height in the solar atmosphere \citep{uitenbroek06, carlos17}. 
 
Multi-line polarimetry allows us to distinguish anisotropies and to diagnose their three dimensional structures in the solar atmosphere.
\citet{hector07} derived magnetic structures extending from the photosphere into the chromosphere from full Stokes profiles in chromospheric Ca\,\emissiontype{II} lines and photospheric Fe\,\emissiontype{I} lines. 
Furthermore, polarimetry in multiple lines with different Land${\rm \acute{e}}$ factor will help us to determine the fraction of area in which magnetic flux is embedded in a spatial resolution element \citep{stenflo73}, and it with different ratio of the Land${\rm \acute{e}}$ factor to the lifetime of the excited state of a line will help us to determine a turbulent magnetic field within a spatial resolution element \citep{stenflo82}.% or a turbulent magnetic field \citep{stenflo82}.}

%飛翔体について

A few solar spectro-polarimeters that obtain full Stokes spectra in multi spectral windows simultaneously are currently in operational in the world.
Two examples are the MulTi-Raies (MTR) instrument of T${\rm \acute{e}}$l${\rm \acute{e}}$scope Heliographique pour l'${\rm \acute{E}}$ tude du Magnetisme et des Instabilit${\rm \acute{e}}$s Solaires (THEMIS) \citep{ariste00} and Spectro-Polarimeter for Infrared and Optical Regions (SPINOR) of the Dunn Solar Telescope (Socas-Navarro et al. \yearcite{navarro06}, \yearcite{navarro11}).
The MTR and SPINOR can obtain up to 4 different spectral windows within 514 - 866 nm and 525 - 854 nm, respectively.
SPINOR extended the wavelength range up to 1600 nm using an infrared camera that was shared with other instruments \citep{hector05, navarro06}.
%They are mainly used to obtain full Stokes spectra in Fe \emissiontype{I} 630 nm, Ca \emissiontype{II} 849 nm, Ca \emissiontype{II} 854 nm, and Ca \emissiontype{II} 866 nm \citep[e.g. ]{ariste01, navarro05}.

The Domeless Solar Telescope (DST) at Hida observatory, Kyoto University, is a 60 cm aperture Newton-Gregorian type telescope equipped with two Czerny-Turnar spectrographs \citep{nakai85}.
One is a vertical vacuum spectrograph (VS), which enables us to observe a single spectral region with a high dispersion, and the other is a horizontal spectrograph (HS), which enable us to make simultaneous multi-line spectrometry in as many wavelength windows as the number of cameras.

Polarimeters were developed on the VS of the DST so far.
\citet{makita91} developed a polarization modulator and an analyzer composed of a rotatable quarter waveplate and a Wallaston prism, and they measured instrumental polarization of the DST at 630 nm. 
Because the optical configuration of DST is not axisymmetric and details of the coating of mirrors of the DST are not known, it is important to measure the instrumental polarization of the DST to realize accurate polarization measurement.
\citet{kiyohara04} replaced the quarter waveplate with a waveplate whose retardation is 126.7$^{\circ}$ at 630 nm.
In addition, with plastic sheet polarizers attached to the entrance window of the DST, they measured instrumental cross-talks among Stokes ${\it Q}$, ${\it U}$, and ${\it V}$ in 630 nm, and modeled the instrumental polarization with Mueller matrices of a two-oblique-mirror system.
On the other hand, \citet{hanaoka09} developed a ferroelectric liquid crystal polarization modulator, and calibrated the instrumental polarization of the DST in 590 nm, 630 nm, and 656 nm.
Finally, \citet{anan12} equipped the polarimeter \citep{kiyohara04} with a Super-Achromatic True Zero-Order Waveplates (APSAW, manufactured by Astropribor Co.), and calibrated the instrumental polarization in a range of the wavelength from 400 nm to 1100 nm.

In order to accurately determine the magnetic field and other physical quantities, and their structures in the solar atmosphere, we developed a system on the HS of the DST that enables us to observe full Stokes spectra in multi-wavelength windows simultaneously within wide wavelength range covering from 500 nm to 1100 nm.
%in a photometric sensitivity of $3 \times 10^{-4}$ and an accuracy of 0.08
Here, an axis which rotates $45^{\circ}$ from positive $Q$ in counterclockwise looking at the source is defined as positive $U$, and the right circular polarization, when an electric vector rotates clockwise looking at the source, is defined as positive $V$.
Our counting system for the matrix coordinates start at 0.
In this paper, we present an overview of the new polarimeter (section \ref{sec.overview}), a spectro-polarimeter for measurement of Mueller matrix of optical components (section \ref{sec.mmsp2}), a polarization modulator and an analyzer (section \ref{sec.pol}), polarization calibration of the new system (section \ref{sec.cal_system}), examples of the observation (section \ref{sec.example}), and summary (section \ref{sec.sum}). 

%%%%%%%%%%%%%%%%%%%%%%%%%%%%%%%%%%%%%%%%%%%%%
\section{Overview of the polarimeter}
\label{sec.overview}

\begin{figure}[htbp]
	\begin{center}
   	\includegraphics[width=160mm]{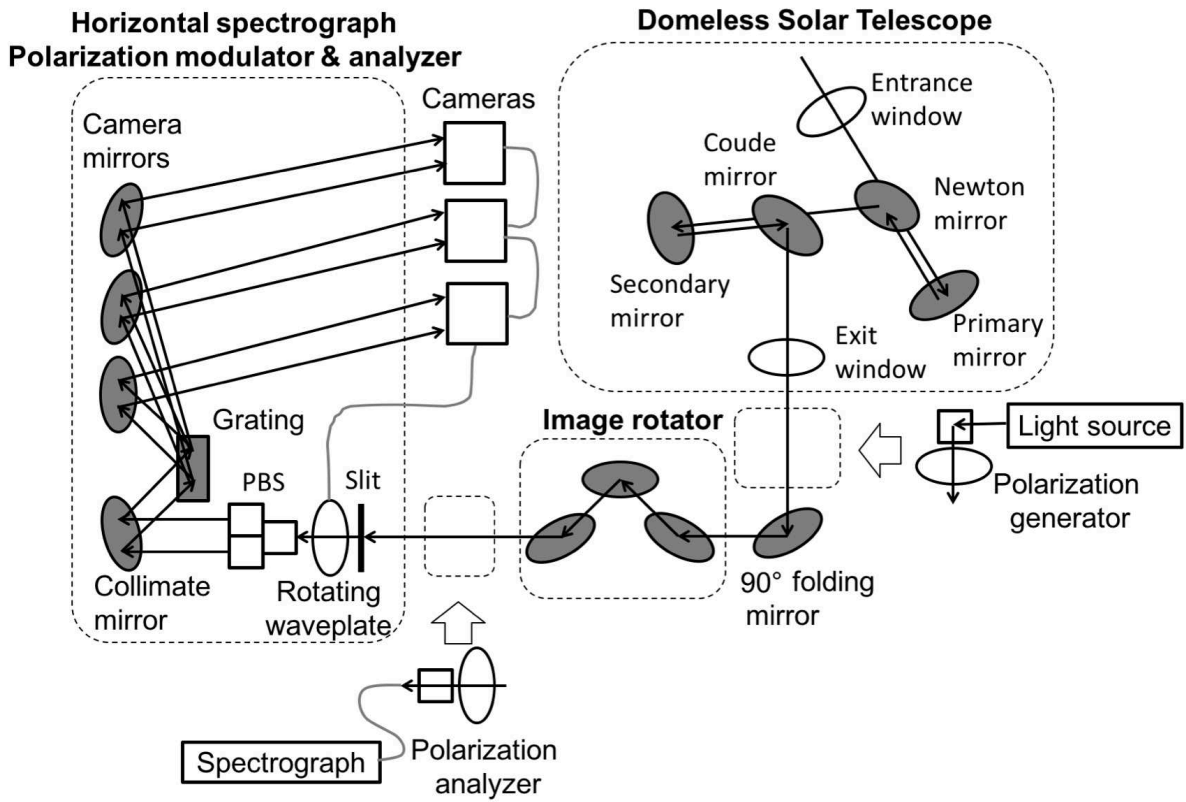}
	\end{center}
	\caption{
	Schematic of the polarimeter.
	Black solid arrows indicate the ray path, and gray lines are trigger cables.
	In fact, the horizontal spectrograph has six camera mirrors.
	Polarization generator and analyzer are used for component calibration.
	}
	\label{fig.overview}
\end{figure}

Figure \ref{fig.overview} displays the overview of the optical system of the polarimeter.
It consists of the DST, an image rotator, a continuously rotating wave plate, a polarizing beam splitter, a high dispersion spectrograph, fast and large format CMOS cameras and an infrared camera. 

The DST is a 60 cm aperture Newton-Gregorian type telescope mounted on a tower of 22 m above ground level.
%Figure \ref{fig.dst_raypath} shows schematic diagram of the ray path and optical components in the telescope.
The DST on alt-azimuth mount tracks the sun by rotating around a horizontal elevation axis and a vertical azimuthal axis.
The primary and the secondary mirrors form a 30 cm solar image at the slit of the spectrograph in observation rooms at the ground level with a composite focal length of 3219 cm.
The optical path between the entrance and the exit windows is evacuated down to 2 mm Hg.
The DST produces a non-negligible instrumental polarization through oblique reflections (incidence angle of 45$^{\circ}$) at the Newton and Coude mirrors.
The polarization property of the DST is well-described by the Mueller matrix as a function of extinction ratios and retardations of these two mirrors, and rotations about the elevation and the azimuthal axes \citep{makita91}.
 
%The DST is equipped with two Czerny-Turnar spectrographs, that is high dispersion vacuum spectrograph and the HS.
%We developed the new polarimeter on the HS.
From the DST, light goes to the slit of the HS through the image rotator, which has a rotor with three flat mirrors  (section \ref{sec.cal_mmsp2}). 
%The length of the spectrograph slit is 20 mm that corresponds to the angular scale of 128 arcsec.
The polarization modulator of the new polarimeter is a continuously rotating waveplate whose retardation is nearly constant around 127$^{\circ}$ in 500 - 1100 nm, and the polarization analyzer is a polarizing beam splitter (PBS) whose diattenuation is sufficiently high in 500 - 1100 nm (section \ref{sec.pol} \& \ref{sec.cal_modulator}).
They have been located closely behind the slit.
%\textcolor{red}{Temperature dependence of the retardation of the waveplate is smaller than 0.03$^{\circ}$/K in 500 - 1100 nm as was known from the measurement.
%Note that it is small enough compared to the tolerance of 3$^{\circ}$ as described in Sec. \ref{sec.pol}.}
In the HS, a grating, which is selectable from three gratings installed in a turret, diffracts the light on six camera mirrors with a focal length of 10 m.
The groove density and the blaze angles of the three gratings are 1200, 1200, 600 grooves/mm, and 17$^{\circ}$27', 26$^{\circ}$45', 48$^{\circ}$55', respectively.
Their sizes of the ruled area are $15.4 \times 20.6\, {\rm cm^2}$.
Each camera mirror forms a spectrum on a corresponding camera port whose length in the dispersion direction is 70 cm.
Thus, we are able to obtain spectra in as many wavelength windows as the number of cameras at the same time in wide spectral ranges covered by the six camera ports. 
%Table \ref{table.lines} shows examples of line sets.
%
%\begin{table}
%\tbl{Line sets }{%
%\begin{tabular}{ccccccc}  
%\hline\noalign{\vskip3pt} 
%  Observation         		& line 1 & line 2 & line 3 & line 4          \\
%\hline\noalign{\vskip3pt} 
%  Normal         		& Fe\,\emissiontype{I} 630 nm  & Na\,\emissiontype{I} 589 nm & Ca\,\emissiontype{II} 854 nm & He\,\emissiontype{I} 1083 nm          \\
%  Prominence    		& He\,\emissiontype{I} 588 nm  & Ca\,\emissiontype{II} 854 nm & Ca\,\emissiontype{II} 866 nm & He\,\emissiontype{I} 1083 nm          \\
%  Photosphere		& Fe\,\emissiontype{I} 525 nm  & Fe\,\emissiontype{I} 630 nm & Fe\,\emissiontype{I} 684 nm & Si\,\emissiontype{I} 1083 nm          \\
%  SUNRISE\emissiontype{III} / SCIP	& K\,\emissiontype{I} 765 nm  & K\,\emissiontype{I} 769 nm & Ca\,\emissiontype{II} 849 nm & Ca\,\emissiontype{II} 854 nm          \\
%[2pt] 
%\hline\noalign{\vskip3pt}
%\label{table.lines} 
%\end{tabular}}
%\end{table}

Fast and large format CMOS cameras (ORCA-Flash4.0, manufactured by HAMAMATSU Photonics K. K.) and an infrared camera (XEVA640, manufactured by Xenics Co.) are hung on rails of the camera ports.
An origin sensor on the rotating waveplate produces a trigger signal for cameras to start a sequence of exposures with a timing accuracy of better than 1 msec.
They obtain simultaneous images of both orthogonal linear polarization states from the PBS.

%%%%%%%%%%%%%%%%%%%%%%%%%%%%%%%%%%%%%%%%%%%%%
\section{Device for component calibration}
\label{sec.mmsp2}

%\citet{ichimoto06}をうまく引用して、最小限の測定原理や測定手法の説明と、装置の説明。
The Mueller Matrix Spectro-Polarimeter (MMSP) is a photopolarimetric system developed to measure the Mueller matrix of optical elements in a manner of \citet{ichimoto06}. It is a system containing dual rotating waveplates.
Polarization is created and analyzed by the combination of a linear polarizer and a rotating waveplate, then the Mueller matrix of an optical element placed in between the two rotating waveplates can be determined.

\begin{figure}[htbp]
	\begin{center}
   	\includegraphics[width=120mm]{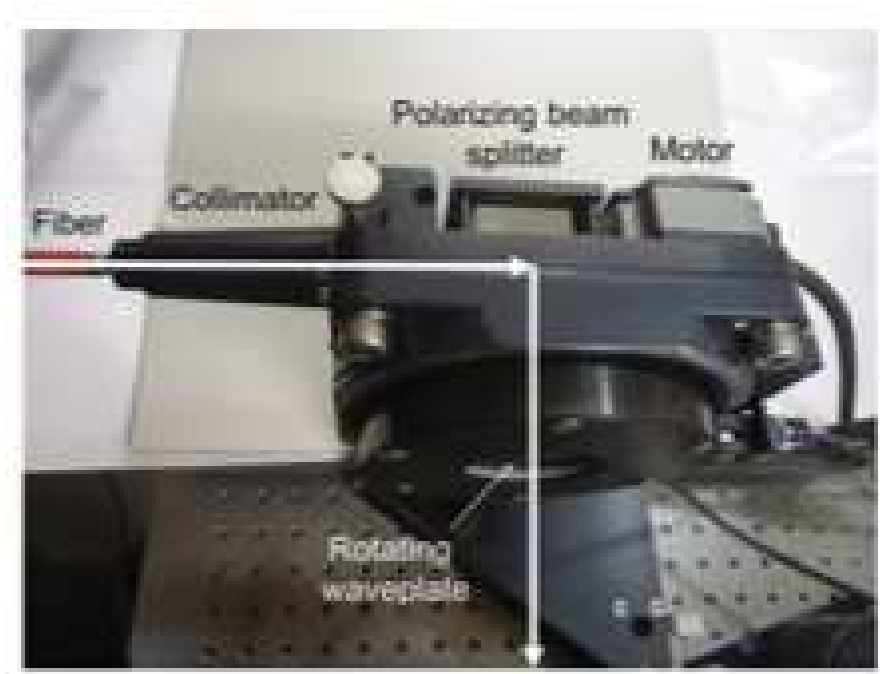}
	\end{center}
	\caption{
	Pictures of the polarization generator unit in the MMSP system.
	The white solid arrows indicate the ray path.
	}
	\label{fig.mmsp2}
\end{figure}
In the polarization generator (figure \ref{fig.mmsp2}), a parallel ray of light created by a collimator and a Tungsten Halogen light source (Ocean Optics, Inc.) is reflected by a polarizing beam splitter so that linearly polarized light is fed to the rotating waveplate.  
The light source provides a continuum spectrum and the polarizing beam splitter has an extinction ratio higher than 300 in a wavelenegth range of 400 - 1100 nm. The angular resolution of the rotating wave plate is 0.02$^{\circ}$. The unit is mounted between the exit window of the DST and the 90$^{\circ}$ folding mirror when the calibration of the imaging rotator and the folding mirror is performed, while the polarization analyzer unit is placed at the exit of the image rotator. The waveplates of the polarization generator and analyzer rotate stepwise by 4.5$^{\circ}$ and 22.5$^{\circ}$, respectively, and spectra covering 400 - 1100 nm are taken at 80 positions in rotation angles of the waveplates to obtain the Mueller matrix.
\section{Polarization modulator and analyzer}
\label{sec.pol}

\begin{figure}[htbp]
	\begin{center}
   	\includegraphics[width=140mm]{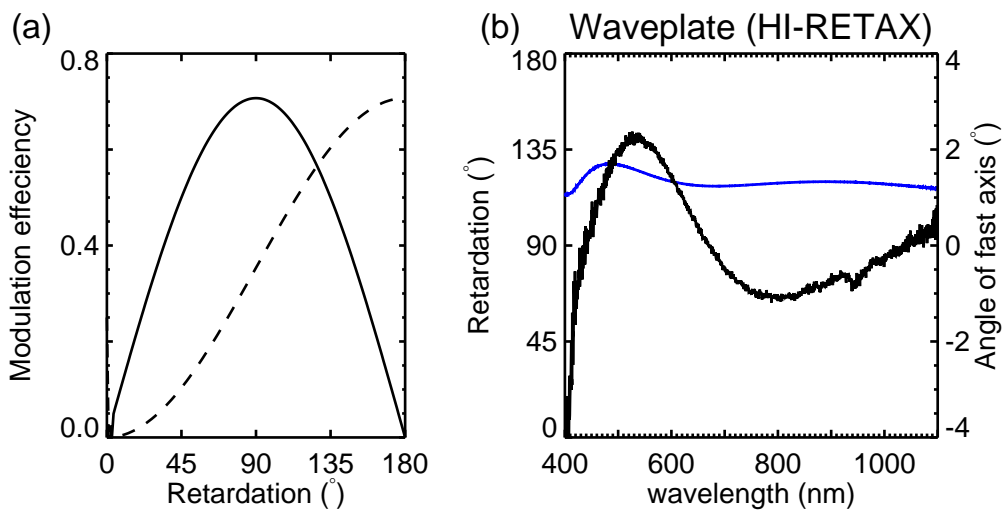}
	\end{center}
	\caption{
	(a) Modulation efficiency \citep{iniesta2000} as a function of the retardation, and (b) retardation (blue) and orientation of the fast axis (black) of the waveplate of the polarization modulator measured by using a photopolarimetric measurement system of Mueller matrix. In the left panel, the solid and the dashed lines indicate the modulation efficiencies for circular and linear polarizations, respectively.
%	
%	\textcolor{blue}{Right: }retardation (blue) and orientation of the fast axis (black) of the waveplate of the polarization modulator measured by using a photopolarimetric measurement system of Mueller matrix.\textcolor{blue}{Left: } modulation efficiency \citep{iniesta2000} as a function of the retardation.
%	In the left panel, the solid and the dashed lines indicate the modulation efficiency for circular and linear polarizations, respectively.
	}
	\label{fig.waveplate}
\end{figure}
\begin{figure}[htbp]
	\begin{center}
   	\includegraphics[width=70mm]{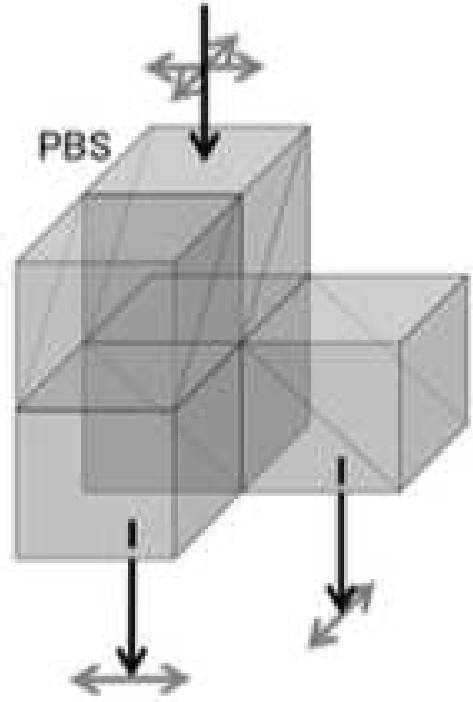}
	\end{center}
	\caption{
	Schematic of the PBS unit.
	Black and gray arrows indicate rays of light and orientation of the polarization, respectively.
	}
	\label{fig.pbs_1}
\end{figure}
The polarization modulator is a rotating waveplate.
The waveplate is super achromatic with the Pancharatnam configuration \citep{pancharatnam55a, pancharatnam55b}.
It has 30 mm diameter and consists of five layers of birefringent polymer film (HI-RETAX, LUCEO Co., Ltd) sandwiched by glass plates (BK7) with anti-reflection coating on it.
%We used a HI-RETAX manufactured by LUCEO Co., Ltd. with a diameter of 30 mm as the modulator .
%The HI-RETAX was assembled from five birefringent wavelength films in Pancharatnam configuration \citep{pancharatnam55a, pancharatnam55b}, and held between aluminum coated glasses (BK7) to decrease internal reflection.
Figure \ref{fig.waveplate}(a) shows modulation efficiency \citep{iniesta2000} as a function of the retardation, and (b) shows the retardation of the waveplate and the orientation of its axis as a function of wavelength. 
They demonstrate that the waveplate has high modulation efficiencies for Stokes {\it Q}, {\it U}, and {\it V} in wavelength range covering from 500 nm to 1100 nm.
Variations of the retardation in operating temperatures, 22 - 28 $^{\circ}$C, are much smaller than requirement values, $\pm 3^{\circ}$, determined through a method described in Sec. 5.1, because its temperature dependence is less than 0.03$^{\circ}$/K in the wavelength range.
Orientation of the fast axis of the waveplate changes with the wavelength in a range of $\pm 2^{\circ}$.
A stepping motor (CRK523PAP-N5, manufactured by ORIENTAL MOTOR  Co., Ltd.) rotates the waveplate with a period longer than 0.3 s. 
An origin sensor (EE-SPX742, manufactured by OMRON Co.) produces a trigger signal for the camera to start a sequence of exposures. 

We used the PBS manufactured by SIGMA KOKI Co., Ltd. as the polarization analyzer.
A dielectric multilayer coating on a diagonal surface inside the PBS splits the incident light into orthogonal linear polarizations with a extinction ratio of 1:300 or higher in the wavelength range covering from 500 nm to 1100 nm.
Figure \ref{fig.pbs_1} shows the PBS unit, which consists of the PBS and four cube prisms.
%They were made of glass (S-TIM35).
Entrance and exit surfaces have anti-reflection coating to decrease internal reflection.
Optical path lengths of both rays are equal \citep{collados07}, but there are observed small deflections of the existing beams due to a manufacturing error.
We applied additional wedge plates at the exit of two beams to reduce the angular deviations.
%of both lights are 18 arcmin and 29 arcmin due to .
%Then, we reduced the angular deviation by using two wedges.
%

\begin{figure}[htbp]
	\begin{center}
   	\includegraphics[width=140mm]{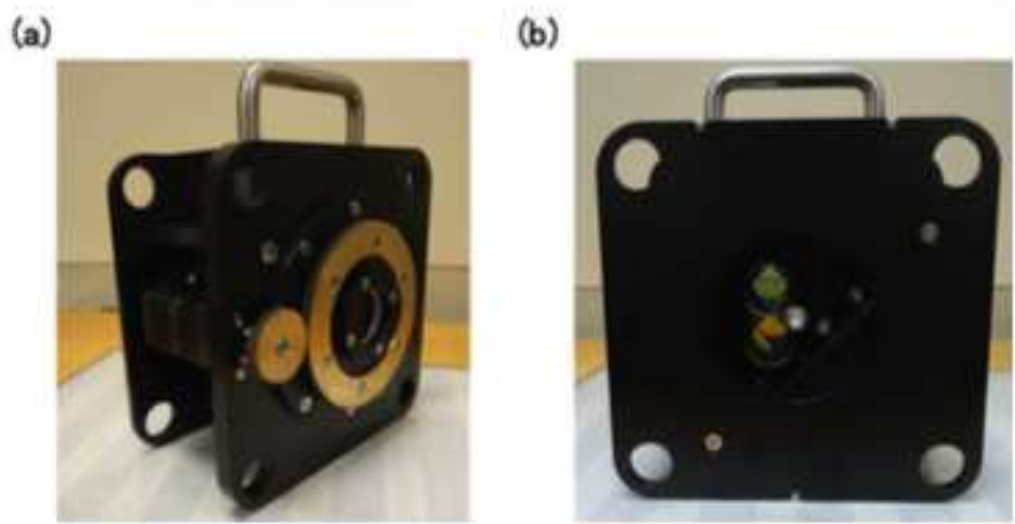}
	\end{center}
	\caption{
	Pictures of the polarization modulator and analyzer unit composed of the rotating waveplate, the PBS and the wedges. Front side (a) back side(b).
	}
	\label{fig.pic_pmu_pbs}
\end{figure}
Figure \ref{fig.pic_pmu_pbs} shows the unit composed of the polarization modulator and the analyzer.
The waveplate is attached to a rotating part held by a pressurized bearing, and rotates with an angular resolution of 0.09$^{\circ}$.
%Reducing the rotating speed using the gears by 15 times increases the accuracy of the rotation speed of the waveplate up to 0.006$^{\circ}$. %R1で5の設定なので、ステップ角度は0.09度
%Compensating for twist of the rotating part, and for deformation due to the gravity and heat, we designed the unit with a wobbling accuracy less than 0.5$^{\circ}$ and with the eccentricity of $\sim0.01$ mm.
The PBS is mounted in the frame of the unit without a tight fixture to avoid mechanical stress as much as possible.
%Moreover, we designed it as eigenfrequency is not devisable by a power frequency of 60 Hz and the rotating frequency.
%These accuracies are guaranteed at every attachment of the unit to the place behind the slit of the spectrograph.

\begin{figure}[htbp]
	\begin{center}
   	\includegraphics[width=140mm]{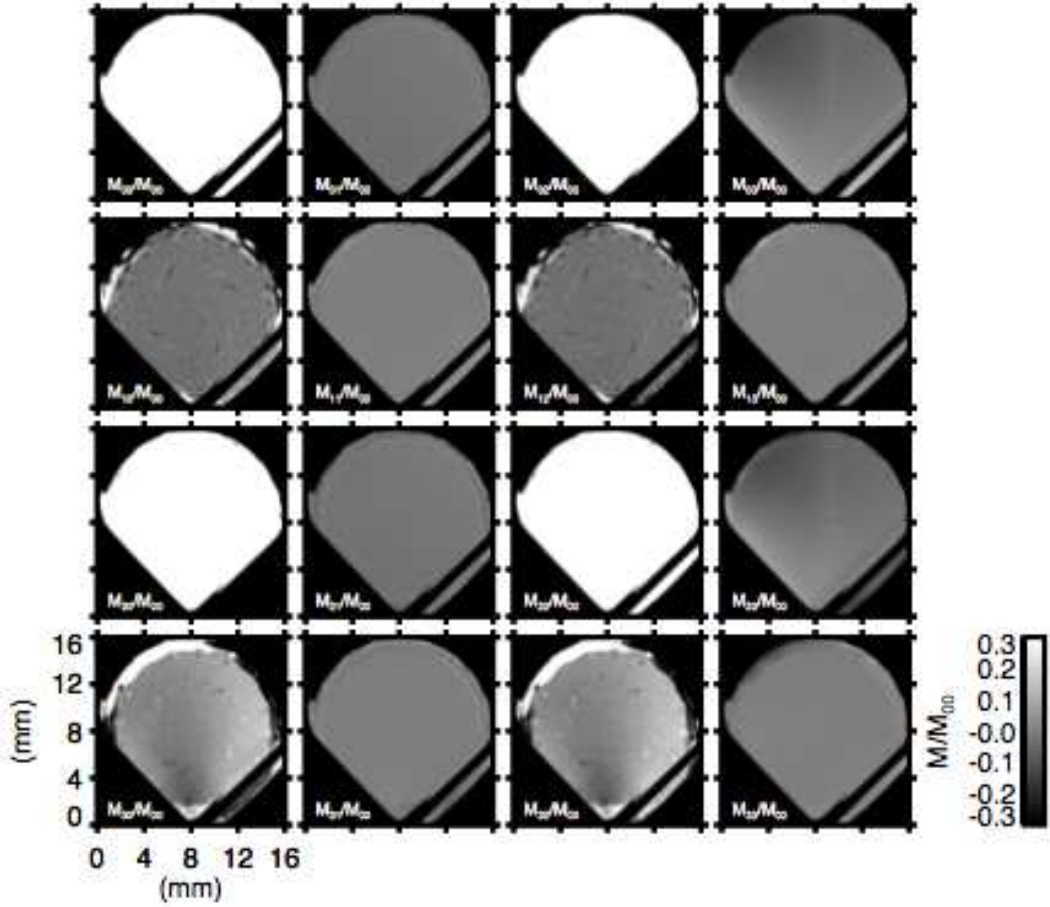}
	\end{center}
	\caption{
	Spatial distributions of the $4\times4$ Mueller matrix elements of a part of the PBS unit in 700 nm.
	}
	\label{fig.pbs_2}
\end{figure}

Figure \ref{fig.pbs_2} shows spatial distributions of the $4\times4$ Mueller matrix elements of a part of the PBS unit at 700 nm.
They were measured with a MMSP (section \ref{sec.mmsp2}) in which we used a CCD camera and a wavelength band path filter in place of the spectrograph.
In the setup of the measurement, the orientations of the orthogonal polarizations of the PBS are in $\pm {\it U}$ directions, respectively.
Spiral patterns in some Mueller matrix elements result from shadows of dusts on the rotating waveplates in the MMSP system.
In addition to diattenuations of typical linear polarizers, it has small ${\it I} \leftrightarrow{\it V}$, and ${\it U} \leftrightarrow {\it V}$ cross-talk.
We interpreted them that the cubes have linear retardations with a fast axis of their diagonal lines ($+$/$-$ ${\it Q}$ directions).
The Mueller matrix of the PBS can be writtem as %which mechanical stresses of binding the cubes cause.
\[
\left(
\begin{array}{cccc}
1 & 0 & 0                  & 0 \\
0 & 1 & 0                  & 0 \\
0 & 0 & \cos{\delta''} & \sin{\delta''} \\
0 & 0 & -\sin{\delta''} & \cos{\delta''}  \\
\end{array}
\right)
\left(
\begin{array}{cccc}
0.5       & 0 & \pm 0.5 & 0 \\
0           & 0 & 0       & 0 \\
\pm 0.5 & 0 & 0.5    & 0 \\
0       & 0 & 0       & 0  \\
\end{array}
\right)
\left(
\begin{array}{cccc}
1 & 0 & 0                  & 0 \\
0 & 1 & 0                  & 0 \\
0 & 0 & \cos{\delta'} & \sin{\delta'} \\
0 & 0 & -\sin{\delta'} & \cos{\delta'}  \\
\end{array}
\right)
\]
\begin{equation}
=
\frac{1}{2}
\left(
\begin{array}{cccc}
1                        & 0 & \pm \cos{\delta'}              & \pm \sin{\delta'}  \\
0                        & 0 & 0                                      & 0 \\
\pm \cos{\delta''} & 0 & \cos{\delta'} \cos{\delta''}  & \sin{\delta'} \cos{\delta''} \\
\mp \sin{\delta''} & 0 & - \cos{\delta'}\sin{\delta''}  & - \sin{\delta'} \sin{\delta''} \\
\end{array}
\right),
\end{equation} 
%確認した
where $\delta'$ and $\delta''$ are the retardations of the cubes before and after the dielectric multilayer coating, respectively.%caused by the stresses.

The measured intensity of spectrum, ${\it I_{\pm}^{obs}}$, in the orthogonal polarizations can be written with the following forms.
%確認した
\begin{eqnarray}
I_{\pm}^{obs} (\theta'_{i}) = \pm \frac{1}{2}R_{\pm} [\, 
                                                           &\pm& C_{1} {\it I^{in}} + C_{2} {\it Q^{in}} + C_{3} {\it V^{in}} 
                                                           + (C_{4} {\it V^{in}} + C_{5} {\it Q^{in}}) \sin{2\theta'_{i}}
                                                           - C_{5} {\it U^{in}} \cos{2\theta'_{i}} \nonumber \\
                                                           &+& C_{6} {\it U^{in}} \sin{4\theta'_{i}}
                                                           + C_{6} {\it Q^{in}} \cos{4\theta'_{i}}
                                                           \,],
\label{eq.modulation1}                                                           
\end{eqnarray}
\[
C_{1} \equiv \gamma,
\]
\[
C_{2} \equiv \gamma \cos{\delta'} \frac{1+\cos{\delta}}{2},
\]
\[
C_{3} \equiv - \gamma \sin{\delta'} \cos{\delta},
\]
\[
C_{4} \equiv - \frac{T}{2\pi} \sin{\frac{2 \pi \gamma}{T}} \cos{\delta'} \sin{\delta},
\]
\[
C_{5} \equiv - \frac{T}{2\pi} \sin{\frac{2 \pi \gamma}{T}} \sin{\delta'} \sin{\delta},
\]
\[
C_{6} \equiv \frac{T}{4\pi} \sin{\frac{4 \pi \gamma}{T}} \cos{\delta'} \frac{ 1 - \cos{\delta} }{2},
\]
\[
\theta'_{i} \equiv \theta_{i} + \frac{\pi \gamma}{T} +\alpha,
\]
where $\pm$ indicate each light of dual orthogonal beams, 
$({\it I^{in}}, {\it Q^{in}}, {\it U^{in}}, {\it V^{in}})$ are the Stokes vector of the incident light on the polarization modulator, 
$\theta_{i}$ is a rotation angle of the waveplate at the {\it i}th exposure start time, 
$R_{\pm}$ is transmittance of the spectrograph for each light, $\gamma$ is the exposure time,  $T$ is the rotation period of the waveplate, $\delta$ is the retardation of the waveplate, and $\alpha$ is the angle between the fast axis of the waveplate and a direction of the PBS when the first exposure is triggered.
It is noted that the $\delta''$ does not affect the modulation of the intensity.
Hereafter, we define the direction of one of the linear polarization of the PBS that tilts 45$^{\circ}$ from the spectrograph slit as the positive $Q^{in}$ on the focal plane of the DST.
%向き入れる？counterclockwise

The orthogonal-polarization spectra are taken simultaneously and sequentially by each camera, while the waveplate rotates continuously.
We derive Stokes ${\it I^{in}}$ from the measurements as
\begin{equation}
{\it I^{in}} = \frac{{\it I_{+}^{obs}} + \frac{R_{+}}{R_{-}} {\it I_{-}^{obs}}}{R_{+} C_{1}}. 
\end{equation}
In addition, the Stokes ${\it Q^{in}}$, ${\it U^{in}}$, and ${\it V^{in}}$ are calculated , respectively, from amplitudes of $\cos{4\theta'}$, $\sin{4\theta'}$, and $\sin{2\theta'}$ in $I_{+}^{obs} - \frac{R_{+}}{R_{-}} I_{-}^{obs}$.  
Determinations of the $\delta$, $\delta'$, $\alpha$, and $R_{+}/R_{-}$ are written in section \ref{sec.cal_modulator}.
Although we are not able to determine the $R_{+}$, it cancels out in reductions of the full Stokes parameters normalized by the Stokes ${\it I}$.
Irregularities or dusts on the surface of the waveplate produce spurious intensity modulation on the camera, but they do not cause a significant error in polarization measurement because the modulation frequency (once/rotation) is different from those from the Stokes ${\it Q}$, ${\it U}$, and ${\it V}$ and the dual beam measurement cancels such in-phase modulations in two channels.

%%%%%%%%%%%%%%%%%%%%%%%%%%%%%%%%%%%%%%%%%%%%%
\section{Polarization calibration}
\label{sec.cal_system}

%%%%%%%%%%%%%%%%%%%%%%%%%%%%%%%%%%%%%%%%%%%%%
\subsection{Requirements for polarimetry}
\label{sec.req}

To define the requirements on the accuracy of the polarization calibration, we take the approach described by \citet{ichimoto08}.
Errors on measured Stokes parameter ${\it S}$ ( $=$ ${\it Q}$, ${\it U}$, and ${\it V}$) can be expressed by 
\begin{equation}
\Delta \biggl( \frac{{\it S}}{{\it I}} \biggr) = \Delta_s \biggl(\frac{{\it S}}{{\it I}}\biggr) + \Delta_b,
\end{equation}
where $\Delta_s$ is the {\it polarimetric accuracy} (scale error), and $\Delta_b$ is the {\it polarimetric sensitivity} (bias error).
The error in the polarization calibration can be described as
\begin{equation}
\Delta \biggl( \frac{S}{I} \biggr) = \left( X_{r}^{-1} X - E \right) \biggl( \frac{S}{I} \biggr),
\end{equation}
where $X$ is the true {\it polarimeter response matrix} \citep{elmore90}, $X_r$ is that used in data reduction, and $E$ is the identity matrix.
We require that the $\Delta_s$ and $\Delta_b$ are smaller than a chosen a scale error $a$ and $\epsilon$ (photometric noise), respectively, then absolute values of $X_{r}^{-1} X - E$ elements should be smaller than those of a tolerance matrix as 
\begin{equation}
\left| X_{r}^{-1} X - E \right| < 
\left(
\begin{array}{cccc}
-                   & a/P_{L}           & a/P_{L}           & a/P_{C}           \\
\epsilon & a                     & \epsilon/P_{L} & \epsilon/P_{C} \\
\epsilon & \epsilon/P_{L} & a                     & \epsilon/P_{C} \\
\epsilon & \epsilon/P_{L} & \epsilon/P_{L} & a                      \\
\end{array}
\right), 
\label{eq.err}
\end{equation}
where $P_{L}$ and $P_{C}$ are expected maximum linear and circular polarization degrees in solar spectra, respectively.

The parameters for chromospheric measurements are different from those of the photospheric measurements.
We adopt $\epsilon = 0.001$, $a = 0.05$, $P_{L} = 0.15$, and $P_{C} = 0.20$ for the photospheric measurement (same as the requirements of {\it Hinode/SOT} \,\cite{ichimoto08}), and $\epsilon = 0.0003$, $a = 0.05$, $P_{L} = 0.01$, and $P_{C} = 0.01$ for the chromospheric measurement.
The tolerance matrix for the chromospheric measurement is
\[
\left(
\begin{array}{cccc}
-           & 5.000 & 5.000 & 5.000 \\
0.0003 & 0.050 & 0.030 & 0.030 \\
0.0003 & 0.030 & 0.050 & 0.030 \\
0.0003 & 0.030 & 0.030 & 0.050                                \\
\end{array}
\right).
\]
Since the elements of the tolerance matrix of the polarimetry can be minimum of those for both cases, the matrix becomes
\[
\left(
\begin{array}{cccc}
-           & 0.333 & 0.333 & 0.250 \\
0.0003 & 0.050 & 0.007 & 0.005 \\
0.0003 & 0.007 & 0.050 & 0.005 \\
0.0003 & 0.007 & 0.007 & 0.050 \\
\end{array}
\right).
\]
We derived tolerances for physical properties of the instrument using equation (\ref{eq.err}).
For example, the tolerance of the waveplate angle is determined as the angle error induced in $X_r$ at which one of the matrix elements violate the equation (\ref{eq.err}).
In this case, the matrix element representing the cross-talk between Stokes $Q$ and $U$ limits the error to be 0.3$^{\circ}$.

%%%%%%%%%%%%%%%%%%%%%%%%%%%%%%%%%%%%%%%%%%%%%
\subsection{Polarization modulator and analyzer}
\label{sec.cal_modulator}

To determine the $\alpha$, $\delta$, $\delta'$, and $R_{+}/R_{-}$, we induced linear polarization oriented to positive ${\it Q^{in}}$ with a polarization degree of $p$ into the polarization modulator.
The modulation equation (equation \ref{eq.modulation1}) is reduced to
\begin{equation}
I_{\pm}^{obs} (\theta'_{i})= \pm \frac{1}{2}R_{\pm} I^{in} [ 
                                                           \pm C_{1} + p C_{2}
                                                           + p C_{5} \sin{2\theta'_{i}}
                                                           + p C_{6} \cos{4\theta'_{i}}
                                                           ].
\label{eq.modulation2}                                                           
\end{equation}
We derived $\alpha$ from the phase of the measured intensity modulation.
We also determine the $\delta$, $\delta'$, and $R_{+}/R_{-}$ from the following expressions
\begin{equation}
\delta = \arccos{
\left\{
\frac{ \frac{T}{4 \pi \epsilon} (C_{+} A_{4\,-} + C_{-}A_{4\,+}) \sin{\left( \frac{4 \pi \epsilon}{T} \right)} - 2 A_{4\,+} A_{4\,-} }
       { \frac{T}{4 \pi \epsilon} (C_{+} A_{4\,-} + C_{-}A_{4\,+}) \sin{\left( \frac{4 \pi \epsilon}{T} \right)} + 2 A_{4\,+}A_{4\,-} }
\right\}
}
\end{equation}
\begin{equation}
\delta' = \arctan{
\left\{
\cos{\left (\frac{2 \pi \epsilon}{T} \right)}
\frac{\cos{\delta}-1}{2 \sin{\delta}}
\frac{A_{2\,+}}{A_{4\,+}}
\right\}
}
\end{equation}
\begin{equation}
\frac{R_{+}}{R_{-}} = - \frac{A_{4\,+}}{A_{4\,-}},
\end{equation}
where, $C_{\pm}$ is constant given by $\frac{1}{2}R_{\pm} I^{in} ( C_{1} \pm p C_{2})$
, $A_{2 \, \pm}$ and $A_{4 \, \pm}$ are amplitudes at a frequency and phase of $\sin{2\theta'_{i}}$ and $\cos{4\theta'_{i}}$, respectively.
%$A_{2 \, \pm} \equiv \pm \frac{1}{2}R_{\pm} I^{in} p C_{5}$
%, and $A_{4 \, \pm}$ is amplitude of $\cos{4\theta'_{i}}$.%, $A_{4 \, \pm} \equiv \pm \frac{1}{2}R_{\pm} I^{in} p C_{6}$.
%Although they are intrinsic characteristics of the waveplate, the PBS, and the spectrograph, we measure them every day when we observe, because they depend on observation environment, for example, room temperature.

We tested the timing accuracy of the synchronization of camera exposures by evaluating the repeatability of $\alpha$ for each camera.
As the results of 30 measurements, the standard deviations of the $\alpha$ for ORCA-Flash4.0 and XEVA640 are equivalent to $0.01^{\circ}$ and $0.3^{\circ}$ of the waveplate angle, respectively.
They are equal to or less than our tolerance of the $\alpha$, $0.3^{\circ}$.% (Sec. \ref{sec.req}).
%人口光源でやったけど、赤外カメラはうまくいかなかった、なぜか？
%2016.12.2の太陽光でのデータを使用して導出した。
%restore,'/sp_pub/save/20161202/camera01/ref.sav' & print,stddev(offsets*!radeg)
%Although the repeatability of the $\alpha$ in 1083 nm with a XEVA640 of $0.3^{\circ}$, is larger than our tolerance, it corresponds that the polarimetric sensitivity of ${\it Q}/{\it I}$ and ${\it U}/{\it I}$ are approximately 0.025 $\times \sqrt{{\it Q}^2 + {\it U}^2}/{\it I}$.

Because the $\alpha$ may have temperature dependences, we need to measure the $\alpha$ before and after the solar observation.

%%%%%%%%%%%%%%%%%%%%%%%%%%%%%%%%%%%%%%%%%%%%%
\subsection{Image rotator}
\label{sec.cal_mmsp2}

\begin{figure}
  \begin{center}
    \includegraphics[width=170mm]{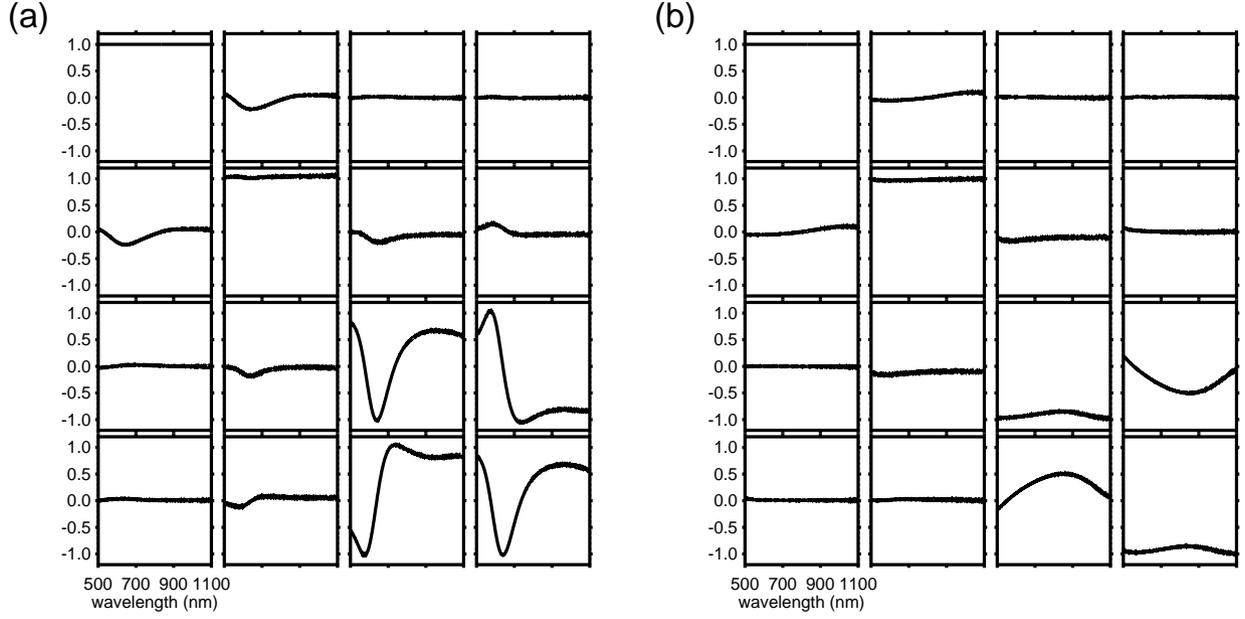}
  \end{center}
  \caption{Mueller matrix of (a) the image rotator at the rotation angle of zero and (b) the $90^{\circ}$ folding mirror as functions of wavelength.
  }
  \label{fig.M_ir}
\end{figure}
%
%\begin{figure}
%  \begin{center}
%   \includegraphics[width=120mm]{./figure/imagerotator.eps}
%   \includegraphics[width=120mm]{./figure/mirror.eps}
%  \end{center}
%  \caption{Configuration of the measurement of image rotator. Polarization is generated by PL2+WP2 and analyzed by WP1+PL1 after going through image rotator. The image rotator can rotate in the plane perpendicular to the light. A beam splitter is not shown in this figure.
%    \caption{Configuration of the measurement of $90^{\circ}$ reflecting mirror. Polarization is generated by PL2+WP2 and analyzed by WP1+PL1 after being reflected by the mirror. The square beyond PL2 is a beam splitter which reflect the light from light source into PL2.}
%      }
%  \label{fig.imagerotator}
%\end{figure}

Compared to the imaging system of the VS, whose polarization character was well modeled with a Mueller matrix  \citep{anan12}, the polarimeter on the HS has additional optics, i.e., a $90^{\circ}$ folding mirror and the image rotator. In this section, we present the Mueller matrices measured with the MMSP system of these two components.

The image rotator consists of three aluminized mirrors with a protection coating on them.
We characterized the image rotator as one Mueller matrix, ${\rm {\bf M}_{IR} } (\theta_{{\rm IR}})$, where $\theta_{{\rm IR}}$ is the rotation angle of the image rotator with $\theta_{{\rm IR}} = 0$ being the position in which the surface of the second mirror is horizontal. The $ {\rm {\bf M}_{IR} } (\theta_{{\rm IR}})$ can be written as:
\begin{equation} 
 {\rm {\bf M}_{IR} } (\theta_{{\rm IR}}) = {\rm {\bf R}} (\theta_{{\rm IR}})  
 						{\rm {\bf M}_{IR} } (0^{\circ}) 
						{\rm {\bf R}} (-\theta_{{\rm IR}}),
\label{eq.model1} 
\end{equation}
where $R(\theta)$ is the rotation matrix in the Q-U plane:
\begin{equation}
  R(\theta) = \left(
    \begin{array}{cccc}
      1 & 0 & 0 & 0 \\
      0 & \cos{2\theta} & \sin{2\theta}  & 0 \\
      0 & -\sin{2\theta}  & \cos{2\theta}  & 0 \\
      0 & 0 & 0 & 1
    \end{array}
  \right).
\end{equation}
%
%and $M_{ir}'$ is %a normalized 4 by 4 matrix independent of $\theta$. "normalized" here means that the first elements ${M_{ir}}_{11}$ is 1.
%a 4 by 4 matrix with the $\theta$ of 0 normalized by the first element of ${M_{ir}}$, that is the (1,1) element ${M_{ir}}_{11}$ (Fig. \ref{fig.M_ir} a).

We measured the Mueller matrix of the image rotator every $2^{\circ}$ sampling from $-60^{\circ}$ to $150^{\circ}$. % \textcolor{red}{using a configuration described in figure \ref{fig.imagerotator}}. 
Figure \ref{fig.M_ir}(a) shows the ${\rm {\bf M}_{IR} } (0^{\circ})$ in the frame where $+{\it Q}$ is in parallel to the surface of the second mirror as measured at 587 nm as functions of angle of the image rotator.
The (1,2) and (2,1) components of the matrix swing above and below zero due to differences between reflectivity for perpendicular and parallel polarized light to the incident plane of the mirrors.%$+Q$ and $-Q$}.
The four elements at right bottom show a strong wavelength dependency of the linear retardation with the axis that is parallel to the incident plane of the mirrors.
The four elements at right top and the four at left bottom are not perfectly equal to 0.
It might be due to due to a slight misalignment of the incident ray from perpendicular to the rotating plane of the image rotator, or due to the misalignments of mirrors inside the image rotator.

\begin{figure}
  \begin{center}
    \includegraphics[width=120mm]{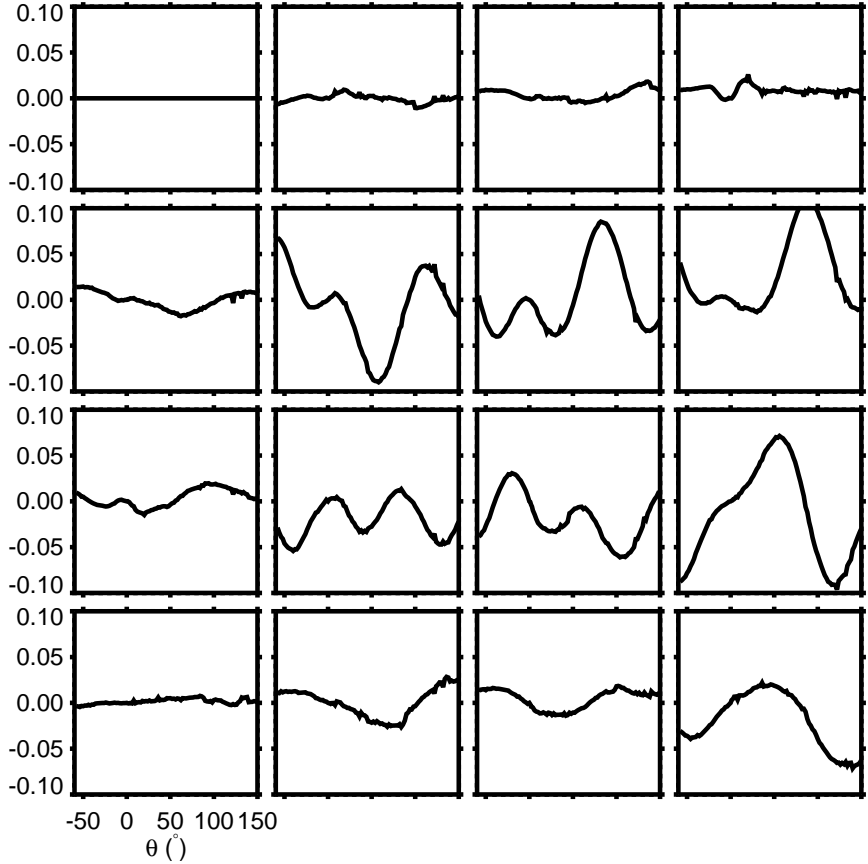}
  \end{center}
  \caption{
  Difference of the measured and inferred Mueller matrices of the image rotator, i.e., ${\rm {\bf M_{IR}} } (\theta_{{\rm IR}})$ and ${\rm {\bf R}} (\theta_{{\rm IR}}) {\rm {\bf M_{IR}} } (0^{\circ}) {\rm {\bf R}} (-\theta_{{\rm IR}})$ at 587 nm as functions of the rotation angle.
    %\textcolor{red}{$R(\theta)M(0)R(-\theta) - M(\theta)$にするべし、別のプロットで0度付近のXrX-Eを見せる。全体キャリブレーションの同じ箇所に注意}
}
  \label{fig.difference}
\end{figure}
%
%\begin{figure}
%  \begin{center}
%    \includegraphics[width=120mm]{./kou20170724/angle(0).png}
%  \end{center}
%  \caption{
%  \textcolor{red}{angle(0).png。eps形式で作る。requirementの横線を引く。}
%  }
%  \label{fig.torelance_of_imagerotator}
%\end{figure}
%
Figure \ref{fig.difference} shows differences between measured ${\rm {\bf M}_{IR} } (\theta_{{\rm IR}})$ and ${\rm {\bf R}} (\theta_{{\rm IR}}) {\rm {\bf M}_{IR} } (0^{\circ}) {\rm {\bf R}} (-\theta_{{\rm IR}})$ at 587 nm. 
%We derive the tolerance matrix of image rotator with $X_r$ being $ R(-\theta) M_{ir, observed}(\theta) R(\theta)$ and $X$ being $M_{ir}'$ \textcolor{red}{これで合ってますか？}.
%Fig. \ref{fig.torelance_of_imagerotator} shows the tolerance matrix where $\theta=1^{\circ}$ as a function of wavelength. \textcolor{red}{測定は2度毎なのに1度のデータはあるのか？}
%Comparing to the requirement in \ref{sec.req}, although the error and standard deviation increase with the wavelength, this result satisfies the requirement of our polarimeter, except (2,1), (3,1) and (4,1) elements. 
%Note that these correspond to the cross-talk from $I$ to $Q$, $U$, $V$, and could be reduced by technique of \citet{kuhn94}.
%Comparing to the requirement of our system derived in Sec. \ref{sec.req}, for instance, elements at row 2 column 3 and row 2 column 4 shows that the residual, as a function of rotation angle, has a maximum reaches 0.1, which is much larger than the requirement 0.007 and 0.005 \textcolor{red}{XrX-Eを計算する。500-1100nmの全波長を考慮する。}.
%Therefore, %since we have measured the Mueller matrix of image rotator every $2^{\circ}$, 
Because the ${\rm {\bf R}} (\theta_{{\rm IR}}) {\rm {\bf M}_{IR} } (0^{\circ})  {\rm {\bf R}} (-\theta_{{\rm IR}})$ does not reproduce the measured ${\rm {\bf M}_{IR} } (\theta_{{\rm IR}})$ perfectly, we use the Mueller matrix of the image rotator at arbitrary angle using the nearest dataset ${\rm {\bf M}_{IR} } (\theta_{{\rm IR}})$ in the calibration of the polarimeter.
For example, ${\rm {\bf M}_{IR} } (5.5^{\circ})$ can be obtained from ${\rm {\bf R}} (-0.5^{\circ}) {\rm {\bf M}_{IR} } (6^{\circ}) {\rm {\bf R}} (0.5^{\circ})$.
We discuss validity of this method in section \ref{sec.mm_total}.
%

%\begin{figure}
% \begin{center}
 %   \includegraphics[width=120mm]{./figure/mirror.eps}
%  \end{center}
%  \caption{Configuration of the measurement of $90^{\circ}$ reflecting mirror. Polarization is generated by PL2+WP2 and analyzed by WP1+PL1 after being reflected by the mirror. The square beyond PL2 is a beam splitter which reflect the light from light source into PL2.}
%  \label{fig.mirror}
%\end{figure}
%
The $90^{\circ}$ folding mirror has silver and protection coating on it and its Mueller matrix, ${\rm {\bf M_{{\rm 90M}}}}$, was also measured by MMSP (figure \ref{fig.M_ir} b). % as in figure \ref{fig.mirror}.
%The Mueller matrix of this reflecting mirror is shown in figure \ref{fig.M_ir} (b).
Note that we also have small (1, 2) and (2, 1) components due to the difference between reflectivity of $+Q$ and $-Q$.
A wavelength-dependent retardation is observed but it is much smaller than that of the image rotator as shown in the four right bottom elements.
The (2, 3) and (3, 2) components show a slight deviation from 0 provably due to a misalignment of the optical setup. 
We redetermine the ${\rm {\bf M_{{\rm 90M}}}}$ in the procedure of the calibration of the entire system.

%
%\begin{figure}
%  \begin{center}
%    \includegraphics[width=120mm]{./figure/M_m.eps}
%  \end{center}
%  \caption{4 by 4 Mueller matrix of $90^{\circ}$ reflecting mirror as a function of wavelength. X-axis is the wavelength from 0.5 $\mu$m to 1.1 $\mu$m}
%  \label{fig.M_m}
%\end{figure}

%%%%%%%%%%%%%%%%%%%%%%%%%%%%%%%%%%%%%%%%%%%%%
\subsection{Entire system}
\label{sec.mm_cal}

\subsubsection{Model}
\label{sec.mm_dst}

Mueller matrices of the DST for the spectropolarimeter on VS were presented by \citet{makita91}, \citet{kiyohara04}, \citet{hanaoka09}, and \citet{anan12}.
%However, they modeled for the polarimetric observation using the VS of the DST.
Here, we present a Mueller matrix, ${\rm {\bf M}_{H}}$, expressing the propagation of the polarization from the entrance of the telescope to the slit of the HS, where the polarization modulator is located closely behind.

We employed the Mueller matrix, ${\rm {\bf T}_{DST}}$, which expresses transformation of the polarization from the DST entrance to the exit window in \citet{anan12}. %, to develop the ${\rm {\bf M_{H}}}$.
In the ${\rm {\bf T}_{DST}}$, unknown parameters are diattenuations of the Newton mirror, $p_{N}$, and of the Coude mirror, $p_{C}$, retardations of the Newton mirror, $\tau_{N}$, and of the Coude mirror, $\tau_{C}$, and unpolarized stray light, $s$, while known parameters are the hour angle, $\theta_{HA}$, and the zenith distance, $\theta_{ZD}$, of the sun.
Direction of positive ${\it Q}$ of the incident light is defined as the east-west direction of the celestial sphere.

The Mueller matrix of the DST for the polarimetric observation using the HS, ${\rm {\bf M}_{H} }$, can be written as
%\begin{eqnarray}
%{\rm {\bf M_{H} } }  =  &&{\rm {\bf R}} (\theta_{ref}) \,\,%{\rm {\bf R}} (\theta_{{\rm offset}}) \,\,
%                                  {\rm {\bf M_{IR}} } (\theta_{IR}) \,\,
%                                  %{\rm {\bf R}} (-\theta_{{\rm offset}}) \,\,
%                                  {\rm {\bf M_{90mir}}} \,\,
%                                  {\rm {\bf R}} (- \theta_{AZ}) {\rm {\bf R}} (\theta_{M}) \nonumber \\
%                                  &&{\rm {\bf T_{DST}}} (\theta_{HA}, \theta_{ZD}, p_{N},\tau_{N}, p_{C}, \tau_{C}, s ),
%\end{eqnarray}
\begin{equation}
{\rm {\bf M}_{H} }  =  {\rm {\bf R}} (\theta_{ref}) \,\,%{\rm {\bf R}} (\theta_{{\rm offset}}) \,\,
                                  {\rm {\bf M}_{IR} } (\theta_{{\rm IR}}) \,\,
                                  %{\rm {\bf R}} (-\theta_{{\rm offset}}) \,\,
                                  {\rm {\bf M}_{{\rm 90M}}} \,\,
                                  {\rm {\bf R}} (- \theta_{AZ}) {\rm {\bf R}} (\theta_{M}) \nonumber \\
                                  {\rm {\bf T}_{DST}} (\theta_{HA}, \theta_{ZD}, p_{N},\tau_{N}, p_{C}, \tau_{C}, s ),
\end{equation}
where $\theta_{ref}$ and $\theta_{M}$ are angles that the axis of the polarization analyzer of the MMSP in the measurement of the image rotator forms with the axis of the polarimeter, and with a line perpendicular to incident plane of the 90$^{\circ}$ folding mirror, respectively, and they are treated as unknown parameters in the Mueller matrix model.
$\theta_{AZ}$ is the azimuth angle of the sun.
%$$ is an offset of the rotation angle of the image rotator from an angle in the $ {\rm {\bf M_{IR}} }$ database
%azimuthの定義どうする？、正の方向は？
Finally, unknown parameters of the ${\rm {\bf M}_{H}}$ are $\theta_{ref}$, $\theta_{M}$, $p_{N}$, $\tau_{N}$, $p_{C}$, $\tau_{C}$, $s$, and the ${\rm {\bf M}_{90M}} $, while the ${\rm {\bf M}_{IR} } (\theta_{IR})$ is evaluated from the measured value as described in section \ref{sec.cal_mmsp2}. 

%ドームレスのモデルの行列も明記したほうがよいでしょうか？

%\subsubsection{Calibration unit}
%\label{sec.mm_turret}
　

\subsubsection{Results}
\label{sec.mm_total}

\begin{figure}
  \begin{center}
    \includegraphics[width=140mm]{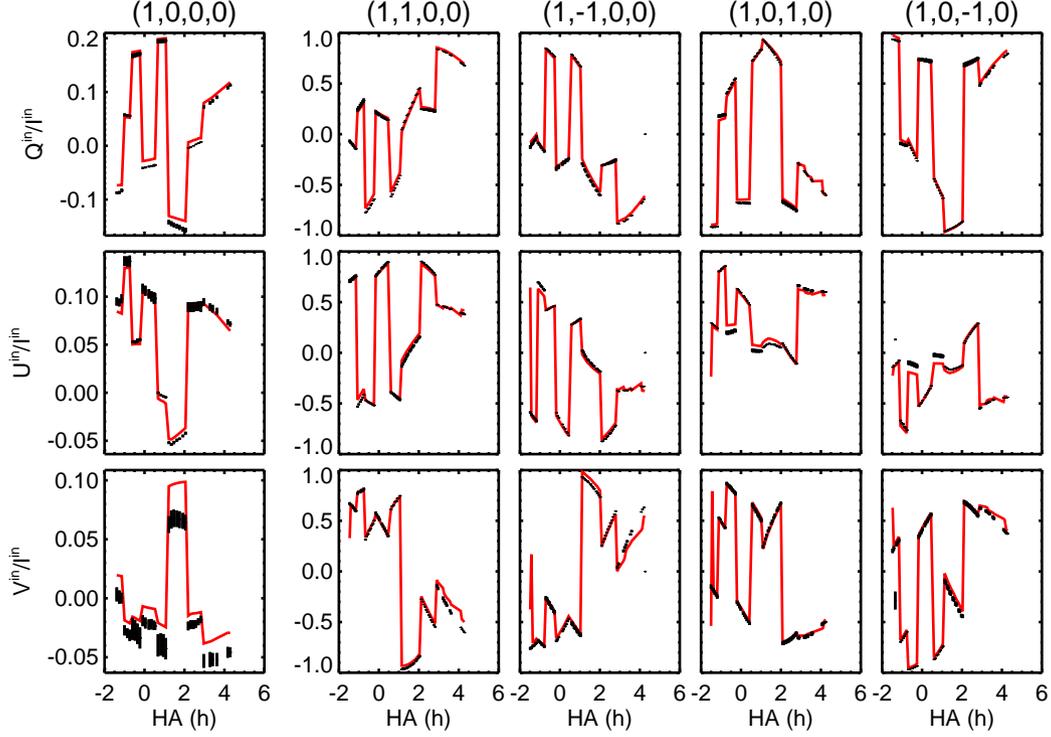}
  \end{center}
  \caption{
  Stokes parameters (${\it Q^{in}}/{\it I^{in}}$, ${\it U^{in}}/{\it I^{in}}$, and ${\it V^{in}}/{\it I^{in}}$) of the incident light into the polarization modulator for unpolarized and linearly polarized incident lights into the entrance window of the DST as functions of hour angle of the sun.
  Data were taken at 589 nm on December 2nd, 2016.
  Black bars are observed values, and red curves are the fitting curves with the DST model.
  The black bars scale the standard deviation of each quantity over the detector.
  The input Stokes vector to the DST are unpolarized, ${\it I} = {\it Q}$, ${\it I} = -{\it Q}$, ${\it I} = {\it U}$, and ${\it I} = -{\it U}$ from the left to the right, as they are indicated at the top of each column.
  The angles of the image rotator were changed at hour angles of \timeform{-1h09m}, \timeform{-0h44m}, \timeform{-0h14m}, \timeform{0h29m}, \timeform{1h04m}, \timeform{2h03m}, and \timeform{2h50m}.
  The minimum and maximum Stokes parameter ranges displayed in the four right columns are -1 and 1, respectively.
  Note that the range of the y-axes of the left column are different from those of the other plots.
  }
  \label{fig.res1}
\end{figure}
\begin{figure}
  \begin{center}
    \includegraphics[width=140mm]{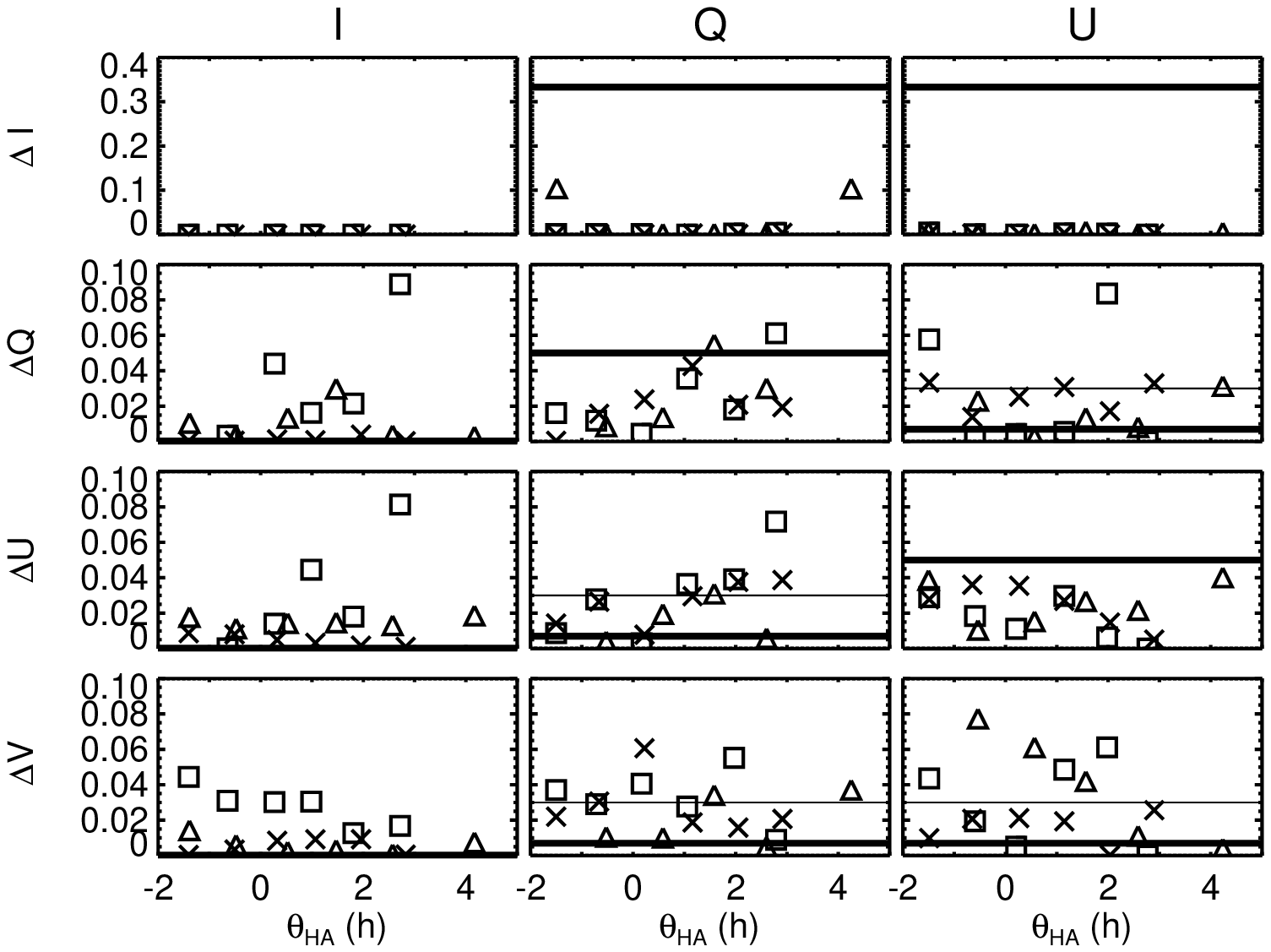}
  \end{center}
  \caption{
	Time evolutions of the error matrix, $\left| X_{r}^{-1} X - E \right|$, in 589 nm ($\triangle$), 849 nm ($\Box$), and 1083 nm ($\times$) on December 2nd, 2016.
	Thick and thin solid lines indicate tolerances for the photospheric observation and for the chromospheric observation, respectively.
	%All elements are required to be less than these tolerances. 
  }
  \label{fig.res2}
\end{figure}

We took the polarization calibration data by pointing the DST equipped with a remotely controllable turret accommodating linear polarizers to quiet region at the solar disk center.% at the entrance window.
The turret enables us to induce unpolarized light or linear polarizations in different orientations from the entrance window \citep{anan12}.
Figure \ref{fig.res1} shows products of the polarimeter (${\it Q}^{in} / {\it I}^{in}$, ${\it U}^{in} / {\it I}^{in}$, and ${\it V}^{in} / {\it I}^{in}$) in continuum for a well known polarized lights (solar light ${\it Q}={\it U}={\it V}=0$, ${\it I } = {\it Q}$, ${\it I } = -{\it Q}$, ${\it I } = {\it U}$, or ${\it I } = - {\it U}$) entering into the entrance window of the DST.
%The continuum light emitted from the quite regions at the solar disk center is an ideal unpolarized light, because its polarization degree is less than $10^{-6}$ \citep{stenflo05}.
The obtained Stokes parameters %were produced by the DST when the ideal unpolarized light enters, or 
are results of transformation by the DST from those of the light that enters the DST.
They change with the hour angle of the sun and the angle of the image rotator, because the transformation depends on the rotation around the horizontal elevation axis and the vertical azimuth axis, and the angle of the image rotator.
Sudden jumps  of the Stokes parameters in figure 9 correspond to intended changes of angle of the image rotator that we applied to sample the wide range of the parameter.
We fit the obtained Stokes parameters by Levenberg-Marquardt least-squares minimization method to determine the unknown parameters of the ${\rm {\bf M_{H}}}$ and the exact angle of the remotely controllable turret. 
The curves of the instrumental polarization by the DST model fairly well reproduces the experimental Stokes parameters.

We estimated an accuracy of the polarization model in reproducing the observed Stokes parameters and the polarimetric sensitivity of our polarimetry from the obtained Stokes parameters and the fitting.
Figure \ref{fig.res2} shows $\left| {\rm {\bf X_{r}}}^{-1} {\rm {\bf X}}' - {\rm {\bf E}} \right|$ in 589 nm, 849 nm, and 1083 nm as a function of hour angle of the sun on December 2nd, 2016, where ${\rm {\bf X_{r}}}$ and ${\rm {\bf X}}'$ are the {\it polarimeter response matrices} inferred from the polarization model and derived from the observed Stokes parameters, respectively.
The accuracy and sensitivity evaluated from equation (\ref{eq.err}) are approximately less than $6\%$ and $10\%$, respectively.
Since we do not have a circular polarizer on the calibration turret, we cannot evaluate the accuracy and sensitivity when circular polarized light enters.
%Figure \ref{fig.res3} shows $\left| {\rm {\bf X_{r}}}^{-1} {\rm {\bf X}}' - {\rm {\bf E}} \right|$ matrix as function of the offset angle of the image rotator from the neighboring angle in the Mueller matrix dataset obtained by the experiment.
Because $\left| {\rm {\bf X_{r}}}^{-1} {\rm {\bf X}}' - {\rm {\bf E}} \right|$ matrix does not depend on the offset angle of the image rotator from the neighboring position in the Mueller matrix dataset obtained by the experiment, the method for calculating the Mueller matrix of the image rotator described in section \ref{sec.cal_mmsp2} is sufficiently reasonable.

Figure 10 tells us that for some elements of the polarimeter response matrix, our polarization model does not allow us to provide a sufficiently accurate value, especially for the first column and off-diagonal elements of the second and third column.
The cross-talk from Stokes ${\it I}$ to ${\it Q}$, ${\it U}$, and ${\it V}$ (the first column) can be corrected by using spurious polarization in the continuum in observational data, because its polarization degree is less than $10^{-6}$ \citep{stenflo05} on the solar disk not very close to the limb.
%Subtracting the polarization degree in the continuum from the Stokes profiles, we can get the polarimetric sensitivity for photospheric ($P_{L}=0.15$) and chromospheric ($P_{L}=1\%$) measurements of $1.2\%$ and $0.08\%$, respectively.
For correcting the remaining cross-talk between the linearly and circularly polarized states, we will estimate them from Stokes spectral profiles of photospheric lines that have sufficiently strong Zeeman signature by assuming that the Stokes ${\it Q}$ and ${\it U}$ profiles are statistically symmetric, and the Stokes ${\it V}$ profiles are statistically anti-symmetric in sunspots.
We determine the cross-talk elements and remove them by using the technique proposed by \citet{kuhn94}.
Finally, after applying all these corrections, the polarimetric error from cross-talk among Stokes parameters is suppressed below the level of random noise of the detectors except the cross-talk between ${\it Q}$ and ${\it U}$ that could be about $\lesssim$8\% or $\lesssim$2.3$^{\circ}$ in the angle of the transversal component of magnetic field.
%th=fingen(90)+!dtor
%q=cos(2th)
%u=sin(2th)
%y0=atan(u,q)*!radeg/2.
%y1=atan(u+0.08,q)*!radeg/2.
%y2=atan(u,q+0.08)*!radeg/2.
%plot,y1-y0
%oplot,y2-y0

%%%%%%%%%%%%%%%%%%%%%%%%%%%%%%%%%%%%%%%%%%%%%
\section{Observation of an active region}
\label{sec.example}

%%%%%%%%%%%%%%%%%%%%%%%%%%%%%%%%%%%%%%%%%%%%%
%\subsection{Active region}
%\label{sec.ex_ar}

%%%%%%%%%%%%%%%%%%%%%%%%%%%%%%%%%%%%
%%%%%%%%%%%%%%%%%%%%%%%%%%%%%%%%%%%%
%%%%%%%%%%%%%%%%%%%%%%%%%%%%%%%%%%%%

\begin{figure}[htbp]
	\begin{center}
   	\includegraphics[width=160mm]{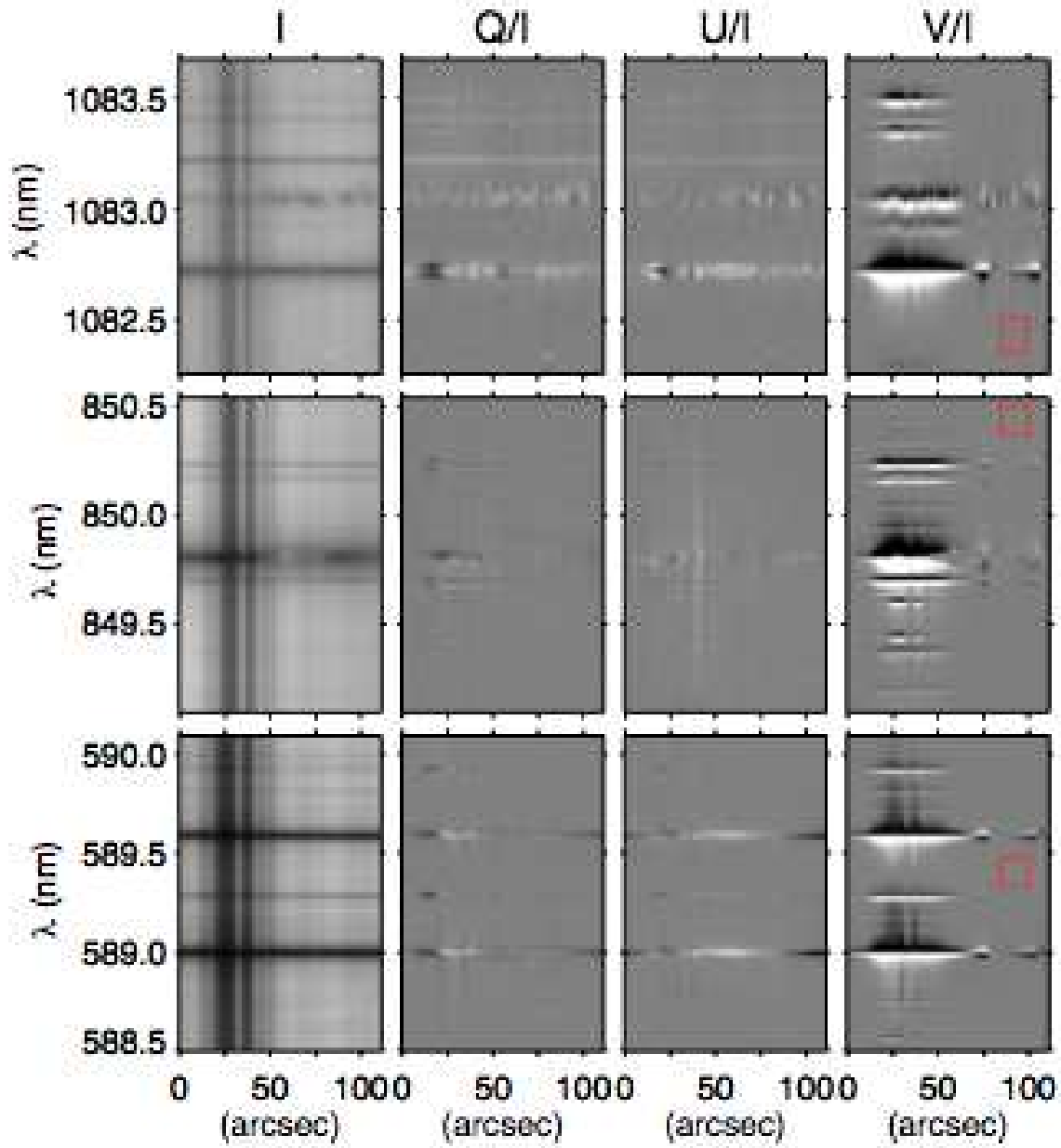}
	\end{center}
	\caption{
	Full Stokes spectra in 589 nm, 849 nm, and 1083 nm (from bottom to top).
	Each column shows Stokes ${\it I}$, ${\it Q}/{\it I}$, ${\it U}/{\it I}$, and ${\it V}/{\it I}$ from left to right.
	The red square boxes mark the regions where we derive noise-to-signal ratios.
	}
	\label{fig.sunspot}
\end{figure}
\begin{figure}[htbp]
	\begin{center}
   	\includegraphics[width=100mm]{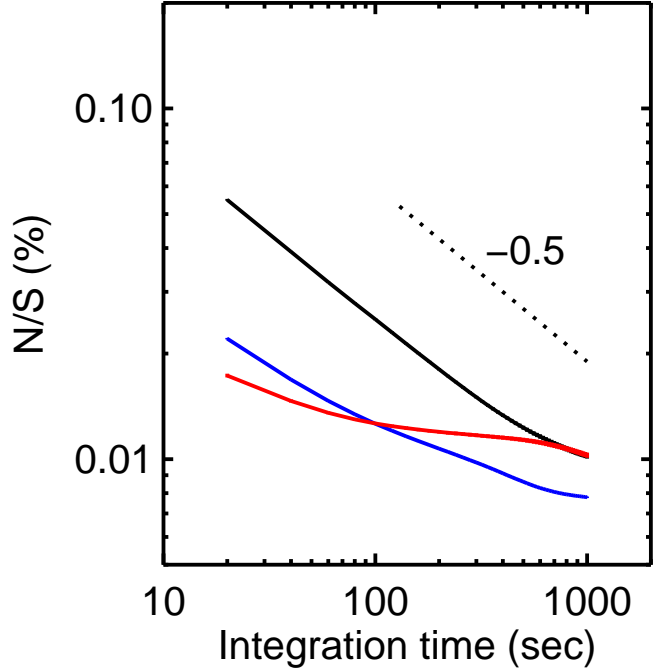}
	\end{center}
	\caption{
	Noise-to-signal ratios (N/S) in 589 nm (blue), 849 nm (black), and 1083 nm (red) as functions of the integration time.
	Black doted line indicates a theoretical expectation in case of the Poisson noise with a power law index of $-0.5$.
	}
	\label{fig.sn_time}
\end{figure}

An active region NOAA 12615 was observed in the Na \emissiontype{I} 589 nm, Ca \emissiontype{II} 849 nm, and He \emissiontype{I} 1083 nm simultaneously using the polarimeter on December 3rd, 2016.
The spectra in 589 nm, 849 nm, and 1083 nm were taken with ORCA-Flash4.0, ORCA-Flash4.0, and XEVA640, respectively.  
The exposure times of all three cameras are $200$ ms, and the spin rate of the rotating waveplate is $0.1$ Hz.
Slow rate of polarization modulation is an issue of our system.
However, 0.1 Hz (10 sec) is about 3 orders of magnitude slower than typical time scale of seeing, so even 10 times higher spin rate will not improve the situation much.
We rather intend to collect sufficient number of photons in a single exposure and suppress the accumulation of the readout noise.
Cross-talks caused by the seeing will be calibrated in post-processing using the unpolarized continuum and Stokes profiles of Zeeman sensitive photospheric lines in observation data.
Figure \ref{fig.sunspot} shows full Stokes spectra in the Na \emissiontype{I} 589 nm, Ca \emissiontype{II} 849 nm, and He \emissiontype{I} 1083 nm composed from 100 frames of spectral images taken in 20 sec.
The slit width is 0.2 mm corresponding to 1.28 arcsec on the solar image.
Spatial samplings along the slit in 589 nm, 849 nm, and 1083 nm are respectively 0.6 arcsec pixel$^{-1}$, 0.6 arcsec pixel$^{-1}$, and 0.5 arcsec pixel$^{-1}$, and spectral samplings are respectively 4.7 pm pixel$^{-1}$, 3.6 pm pixel$^{-1}$ and 4.7 pm pixel$^{-1}$. 

Noise-to-signal ratios of the data are estimated as the standard deviations of the Stokes ${\it V}/{\it I}$ in continuum in a spatial-spectral region indicated by the squares in the figure \ref{fig.sunspot}, where we avoided interference fringes in the selection of the region.
The results are shown in figure \ref{fig.sn_time} as a function of the integration time.
The ratios decrease with the integration time, and reach below $3 \times 10^{-4}$ in 10 sec for 589 nm and 1083 nm, and 60 sec for 849 nm.
Because the slope of the function in 849 nm is nearly equal to $-0.5$, the photon noise is the dominant source of the noise in 849 nm.
With the same reasoning, the photon noise is the dominant source of the noise also in 589 nm up to the noise-to-signal ratio of $0.01$ \%.
On the other hand, the slope of the function of 589 nm below the noise-to-signal ratio of $0.01$ \% and that of 1083 nm are softer than $-0.5$, indicating that the dominant source of error may arise from some other systematic effect.

%\textcolor{red}{赤外カメラの非線形性補正が甘い。理由１、プロミネンスは非線形補正しないほうが良い、理由２、Iで規格化してもIからのクロスートークあり} 
%\textcolor{red}{Naの吸収が深いところにスリット方向のパターンが残る。吸収が強いところはPBSの遅延量の補正がいらないのか？わけ分からん}
%\textcolor{red}{CaのU?にだけ、Vからのクロストークじゃないけど、Vのような信号が一定に存在する。}

%%%%%%%%%%%%%%%%%%%%%%%%%%%%%%%%%%%%%%%%%%%%%
%\subsection{Filament}
%\label{sec.ex_fi}

%観測できたら執筆。

%%%%%%%%%%%%%%%%%%%%%%%%%%%%%%%%%%%%%%%%%%%%%
\section{Summary}
\label{sec.sum}

Multi-lines spectro-polarimetry allows us to diagnose structures of magnetic field vectors along the line of sight, or develop new diagnoses of other vector physical quantities in the solar atmosphere, for examples, electric fields, radiation field and particle stream.
We developed a new spectropolarimeter on the HS of the DST at Hida Observatory.
The system enables us to obtain full Stokes spectra in multi-wavelength windows simultaneously in a range of wavelength covering from 500 nm to 1100 nm.
The cross-talk among the Stokes ${\it Q}$, ${\it U}$ and ${\it V}$ after the calibration of the instrumental polarization based on the Mueller matrix model is of the order of 6\%; for example, 5\% Stokes V signal could produce a cross-talk signal of 0.3\% in Stokes Q.
With further post-processing calibration  using the unpolarized continuum and Stokes profiles of Zeeman sensitive photospheric lines in observational data, we aim to achieve a sensitivity of $3 \times 10^{-4}$ in 20 - 60 sec .

\bigskip
\begin{ack}
This work was supported by JSPS KAKENHI Grant Number 22244013, JP16H01177, and JP15K17609.
We thank all the staff and students of Kwasan and Hida Observatory.
We are also grateful to Mr. J. Maekawa for outstanding work for the maintenance of the Domeless Solar Telescope for 36 years.
\end{ack}


\begin{thebibliography}{}

\bibitem[Anan et al. (2012)]{anan12}
   Anan, T., Ichimoto, K., Oi, A., Kimura, G., Nakatani, Y. \& Ueno S. 2012, \procspie, 8446E, 1CA
\bibitem[Casini (2005)]{casini05}
   Casini, R. 2005, \pra, 71, 2505C
\bibitem[Collados et al. (2007)]{collados07}
   Collados, M., Lagg, A., D${\rm \acute{i}}$az Garc${\rm \acute{i}}$ A., J. J., Hern${\rm \acute{a}}$ndez Su${\rm \acute{a}}$rez, E., L${\rm \acute{o}}$pez L${\rm \acute{o}}$pez, R., P${\rm \acute{a}}$ez Ma${\rm \tilde{n}}$${\rm \acute{a}}$, E. \& Solanki, S. K. 2007, ASPC, 368, 611C
\bibitem[del Toro Iniesta \& Collados (2000)]{iniesta2000}
   del Toro Iniesta, J. \& Collados, M. 2000, \ao, 39, 1637
\bibitem[Elmore (1990)]{elmore90}
   Elmore, D. F. 1990, A polarization calibration technique for the advanced stokes polarimeter. NCAR Technical Note NCAR/TN-355+STR, NCAR, Boulder, Colorado.
\bibitem[Foukal \& Hinata (1991)]{foukal91}
   Foukal, P. \& Hinata, S. 1991, \solphys, 132, 307P
\bibitem[Hanaoka (2009)]{hanaoka09}
   Hanaoka, Y. 1991, \pasj, 61, 357H
\bibitem[Hanle (1924)]{hanle24}
   Hanle, W. 1924, ZPhy, 30, 93H
\bibitem[Ichimoto et al. (2006)]{ichimoto06}
   Ichimoto, K., Shinoda, K., Yamamoto, T. \& Kiyohara, J. 2006, PNAOJ, 9, 11I
\bibitem[Ichimoto et al. (2008)]{ichimoto08}
   Ichimoto, K., Lites, B., Elmore, D., Suematsu, Y., Tsuneta, S., Katsukawa, Y., Shimizu, T., Shine, R., Tarbell, T., Title, A., Kiyohara, J., Shinoda, K., Card, G., Lecinski, A., Streander, K., Nakagiri, M., Miyashita, M., Noguchi, M., Hoffmann, C. \& Cruz, T. 2008, \solphys, 249, 233
\bibitem[Kiyohara et al. (2004)]{kiyohara04}
   Kiyohara, J., Ueno, S., Kitai, R., Kurokawa, H., Makita, M. \& Ichimoto, K. 2004, \procspie, 5492, 1778K
\bibitem[Kuhn et al. (1994)]{kuhn94}
   Kuhn, J. R., Balasubramaniam, K. S. Penn, G. K. M. J., Dombard, A. J. \& Lin, H. 1994, \solphys, 153, 143K
\bibitem[Landi Degl'Innocenti \& Landolfi (2004)]{landi04}
   Landi Degl'Innocenti, E. \& Landolfi, M. 2004, ASSL, 307, L
\bibitem[L${\rm \acute{o}}$pez Ariste et al. (2000)]{ariste00}
   L${\rm \acute{o}}$pez Ariste, A., Rayrole, J. \& Semel, M. 2000, \aaps, 142, 137
\bibitem[Makita et al. (1991)]{makita91}
   Makita, M., Funakoshi, Y. \& Hanaoka, Y. 1991, sopo.work, 198M
\bibitem[Nakai \& Hattori (1985)]{nakai85}
   Nakai, Y. \& Hattori, A. 1985, MmKyo, 36, 385N
\bibitem[Pancharatnam (1955a)]{pancharatnam55a}
   Pancharatnam, S. 1955, Proc. Indian Acad. Sci., 41, Issue 4, 130
\bibitem[Pancharatnam (1955b)]{pancharatnam55b}
   Pancharatnam, S. 1955, Proc. Indian Acad. Sci., 41, Issue 4, 137
\bibitem[Quintero Noda et al. (2017)]{carlos17}
   Quintero Noda, C., Shimizu, T., Katsukawa, Y., de la Cruz Rodr${\rm \acute{i}}$guez, J., Carlsson, M., Anan, T., Oba, T., Ichimoto, K. \& Suematsu, Y. 2017, \mnras, 464, 4534Q
\bibitem[Socas-Navarro \& Elmore (2005)]{hector05}
   Socas-Navarro, H. \& Elmore, D. 2005, \apj, 619L, 195S
\bibitem[Socas-Navarro et al. (2006)]{navarro06}
   Socas-Navarro, H., Elmore, D., Pietarila, A., Darnell, A., Lites, B. W., Tomczyk, S. \& Hegwer, S. 2006, \solphys, 235, 55S
\bibitem[Socas-Navarro (2007)]{hector07}
   Socas-Navarro, H. 2007, \apjs, 169, 439S
\bibitem[Socas-Navarro et al. (2011)]{navarro11}
   Socas-Navarro, H., Elmore, D., AsensioRamos, A. \& Harrington, D. M. 2011, \aap, 531A, 2S
\bibitem[Stark (1913)]{stark13}
   Stark, J. 1913, \nat, 92, 401S
\bibitem[Stenflo (1973)]{stenflo73}
   Stenflo, J. O. 1973, \solphys, 32, 41S
\bibitem[Stenflo (1982)]{stenflo82}
   Stenflo, J. O. 1982, \solphys, 80, 209S
\bibitem[Stenflo (2005)]{stenflo05}
   Stenflo, J. O. 2005, \aap, 429, 713
\bibitem[Stenflo (2017)]{stenflo17}
   Stenflo, J. O. 2017, \ssr, 210, 5S
\bibitem[Uitenbroek (2006)]{uitenbroek06}
   Uitenbroek, H. 2006, \asp, 354, 313U
\bibitem[Zeeman (1897)]{zeeman97}
   Zeeman, P. 1897, \nat, 55, 347Z




%\bibitem[L${\rm \acute{o}}$pez Ariste et al. (2001)]{ariste01}
%  L${\rm \acute{o}}$pez Ariste, A., Socas-Navarro, H. \& Molodij, G. 2001, \apj, 552, 871L
   
   
   
\end{thebibliography}
\end{document}